\documentstyle[12pt,epsf]{article}

\newcommand{\beq}{\begin{equation}}
\newcommand{\eeq}{\end{equation}}

\begin{document}
\title{{Study of $\pm J$ Ising Spin Glasses via Multicanonical
Ensemble\thanks{To appear in the Proceedings of the Sixth Annual
Workshop on Recent Developments in Computer Simulation Studies in Condensed
Matter Physics, 22--26 Feb. 1993, Athens, Georgia.}}}

\author{Tarik Celik$^{1,2}$, Ulrich H.E. Hansmann$^{1,3}$
and Bernd Berg$^{1,3,4}$}

\footnotetext[1]{\it Supercomputer Computations Research Institute,
	 Florida State University,Tallahassee,FL 32306,USA}
\footnotetext[2]{\it Department of Physics Engineering, Hacettepe University,
	   06532 Beytepe, Ankara, Turkey}
\footnotetext[3]{\it Department of Physics,
	 Florida State University, Tallahassee, FL 32306, USA}
\footnotetext[4]{\it Wissenschaftskolleg zu Berlin,
	 Wallotstr. 19, Berlin 33, Germany}
\maketitle
\begin{abstract}
We performed numerical simulations of 2D and 3D
Edwards-Anderson spin glass models by using the recently developed
multicanonical ensemble. Our ergodicity times increase with the
lattice size approximately as $V^3$.
The energy, entropy and other physical quantities
are easily calculable at all temperatures from a single simulation.
Their finite size scalings and the zero temperature limits are
also explored.
\end{abstract}

The low temperature phase of the spin glasses has distinct properties
like broken ergodicity and the absence of self-averaging, which in
turn make their numerical investigation an extremely difficult
task \cite{Bhatt}. For temperatures below a bifurcation point,
the spin glass configuration space is supposed to split off into
a number of valleys which are  separated by high energy barriers.
Due to the exponentially increasing relaxation times \cite{Bi1} encountered
in canonical simulations, tunneling between these many thermodynamics
states becomes almost impossible.
Recent simulations \cite{Cara} of the 3D Edwards-Anderson
model in a magnetic field seem to support the mean field
picture \cite{Paris} rather than the alternative droplet model \cite{Huse},
but it can be argued that equilibrium at sufficiently low temperatures
has not been reached \cite{Fish}.

One of the simplest spin glass
systems is described by the Edwards-Anderson Hamiltonian
\begin{equation}
H\ =\ - \sum_{<ij>} J_{ij} s_i s_j ,
\end{equation}
where the sum includes only the nearest neighbours and the exchange
interactions $J_{ij}= \pm 1$ between the spins $s_i = \pm 1 $ are the
quenched random variables. The constraint $ \sum J_{ij} = 0$
is imposed for each realization.

We present a new approach to spin glass simulations which reduces
the exponential slowing to a power law and enables one to sample
independent ground states in one simulation. This is achieved by
exploiting the multicanonical
ensemble \cite{ours1}.


The multicanonical ensemble  can be defined by weight factors
\begin{equation}
P_M (E)\ =\ \exp \left[ -\beta (E) E + \alpha (E) \right] .
\end{equation}
$\alpha (E)$ and $\beta (E)$ are to be determined such that for the
chosen energy range $E_{\min}\le E\le E_{\max}$ the resulting
multicanonical probability density is approximately flat:
\begin{equation}
P_{mu} (E)\ =\ c_{mu}\ n(E) P_M (E)\ \approx\ \hbox{const}
\end{equation}
where $n(E)$ is the spectral density.
In the present study we take $E_{\max} =0$ ($\beta (E) \equiv 0$ for
$E\ge E_{\max}$) and $E_{\min}=E^0$ the ground state energy of the
considered spin glass realization.
The purpose of the function $\alpha (E)$ is to give $\beta (E)^{-1}$
the interpretation of an effective temperature. This leads to the
recursion relation
\begin{equation}
\alpha (E-4)\ =\ \alpha (E) +
\left[ \beta (E-4) - \beta (E) \right] E
\end{equation}
where $ \alpha(E_{\max}) = 0 $ .
The multicanonical function $\beta (E)$ is obtained  via recursive
multicanonical calculations. One performs simulations with $\beta^n (E)$,
$n=0,1,2,...$, which yield probability densities $P^n (E)$ with
medians $E^n_{\tenrm median}$.  We start off with $n=0$
and $\beta^0 (E) \equiv 0$. The recursion from $n$ to $n+1$ reads
$$ \beta^{n+1} (E)\ =
\left\{
   \begin{array}{ll}
    \beta^n (E)\   \hbox{~for} \ E \ge E^n_{\tenrm median};& \\
\\
    \beta^n (E)\ + \\  0.25\times\ \ln \left[ P^n(E+4)/P^n(E) \right]\\
\qquad   \hbox{for\ \ }  E^n_{\tenrm median} > E \ge E^n_{\min}; &\\
\\
    \beta^{n+1} (E^n_{\min})\   \hbox{~for\ \ }  E < E^n_{\min}\, .&{}
\end{array}
\right.
\eqno(5) $$
The recursion is stopped for $m$ with
$E^{m-1}_{\min}=E^0$ being groundstate.

By weighting with $\exp [-\hat\beta E + \beta (E) E - \alpha (E)]$
canonical expectation values
$ {\cal O} (\hat\beta)\ =\ Z({\hat\beta} )^{-1}
\sum_E {\cal O} (E) n(E) \exp (-\hat\beta E) $, where
$ Z(\hat\beta )\ =\ \sum_E n(E) \exp (-\hat\beta E) $
is the partition function, can be reconstructed for all $\hat\beta$
(the canonical inverse temperature).
The normalization constant $c_{mu}$ in equation (3)
follows from $Z(0)=\sum_E n(E) = 2^N$, where $N$ is the total number of
spin variables. This gives the spectral density and allows to calculate
the free energy as well as the entropy.

First, we  simulated the 2D Ising model
by setting all the exchange interactions to $J_{ij}=+1$ on a $50 \times 50$
lattice. Figure 1 stems from this simulation.
Finite lattice energy density and specific heat results of Ferdinand and Fisher
\cite{Ferd} were well reproduced. With $Z(0)=2^{2500}$ used as input,
we obtained $2.07 \pm 0.22$ for the number of groundstates. After having this
check, we considered the  Edwards-Anderson spin glass.

We performed multicanonical simulations of Edwards-Anderson spin glasses
in 2D on $L^2$ lattices with $L = 4, 12, 24, 48$ and in 3D on $L^3$ lattices
with $L = 4, 6, 8$ and $12 $ \cite{ours2}.
We investigated as many as 32 realizations
(having different configurations of quenched random variables $J_{ij}$ ) for
smallest lattices and 4 realizations for the largest. Following an initial
$2 \times 10^5 $ sweeps, the thermal averages were evaluated over
$4 \times 10^5 $ to $8 \times 10^6 $ iterations depending on lattice size.
2 to 10 recursion steps of (5) were needed to reach to an approximately
flat probability density (3).
\begin{figure}[htb]
\vspace{7cm}
\includegraphics{fig1.ps}
\caption{Ising model magnetic probability density.}
\label{fig:Figure 1}
\end{figure}

Our ergodicity time $\tau^e_L$ is defined as the average number of sweeps
needed to change the energy from $E_{\max}$ to $E_{\min}$ and back. In
Figure 2 the ergodicity time for 2D is plotted versus lattice size $L$ on a
log-log scale. A straight line fit gives $\tau^e_L\ \sim\ L^{4.4 (3)}~ $
sweeps,  which in CPU time corresponds to a slowing down
$\sim V^{3.2 (2)}$. Here an exponential form yielded a completely
unacceptable goodness of fit $Q < 10^{-6}$.

The infinite volume groundstate energy and entropy are estimated from
FSS fits of the form $f^0_L = f^0_{\infty} + c/V $. For 2D these fits are
depicted in Figure 3. Our groundstate energy density estimate
$e^0 = -1.394 \pm 0.007$ and groundstate entropy per spin estimate
$s^0 = 0.081 \pm 0.004$ are all consistent with previous MC \cite{Swend}
and the transfer matrix results \cite{Cheng}. The reported groundstate
entropy value translates into a large number of distinct groundstates,
for instance, implies $1.1 \times 10^{81}$ groundstates on $48^2$ lattice.

\begin{figure}[htb]
\vspace{4.5cm}
\includegraphics{fig2.ps}
\caption{Ergodicity times versus lattice size.}
\label{fig:Figure 2}
\end{figure}

For each realization we also evaluated spin glass susceptibility density
$\chi_q = \left< q^2 \right> / N $ and the Binder parameter
$ B_q = {1\over 2} \left[3 - {\left< q^4 \right> /  \left< q^2
 \right>^2} \right]$ . Our results were consistent with the previous
evaluations \cite{Bhatt}.
\begin{figure}[htb]
\vspace{5.1cm}
\includegraphics{fig3.ps}
\includegraphics{fig3a.ps}
\caption{FSS estimate of the infinite volume groundstate energy
	  and entropy per spin.}
\label{fig:Figure 3}
\end{figure}

For $4^2$ lattice we  calculated the exact results by enumerating
all $2^{16}$  configurations per realization and all quantities
were found in agreement with the multicanonical evaluations.

Next we compared our performance with multicanonical simulation to
that of canonical algorithm. For $12^2$ and $24^2$ lattices we
performed an identical number of sweeps  with
multicanonical and canonical simulations for each realization.
For all realizations the canonical tunneling rate is found smaller than the
multicanonical one. When tunneling is concerned, although it strongly
depends on the realization, an
estimated overall improvement factor
$46 \pm 22$  for multicanonical simulations was observed for $12^2$ lattices.
For $24^2$ lattices  we obtained zero tunneling events for most of
the realizations in canonical simulations and about 10 - 20 tunnelings
for the multicanonical. In view of the ergodicity times we achieved,
this is no surprise. It clearly shows the superiority of the
multicanonical approach.

Dimension is a parameter in our codes. No difficulties were encountered
in our investigation of 3D spin glasses
and results were encouraging
\cite{ours2}.
The groundstate energy density $e^0 = -1.786 \pm 0.003 $ and entropy
per spin $s^0 = 0.046 \pm 0.002 $ are consistent with previous
estimates \cite{Morg}.
For 3D we had a slowing down $\sim V^{3.4 (2)}$.

For each realization
we simulate two replica which differ only in the disordered starting
configurations of spin variables $s_i$. Spin glass order parameter $q$
can be defined as the overlap of the two replica \cite{Bi1}
\begin{equation}
q(\hat\beta )\ = \ {1\over N} \sum_{i=1}^N s^1_i s^2_i\,.
\end{equation}
To visualize the low temperature behaviour, we show in Figure 4 the
spin glass order parameter distribution of one of the realizations on
$8^3$ lattice.
\begin{figure}[htb]
\vspace{4.7cm}
\includegraphics{fig4.ps}
\caption{Spin glass order parameter distribution on $8^3$ lattice.}
\label{fig:Figure 4}
\end{figure}

Below the bifurcation temperature one clearly sees
five configuration space valleys which
are separated by high tunneling barriers. The multicanonical simulation
overcomes these energy barriers by connecting back to the disordered
high temperature states. The realization from which Figure 4 is depicted
had nine tunnelings between the degenerate groundstates.
For the Ising ferromagnet of Figure 1,
the number of tunneling events was 118.

Our results make clear that the multicanonical spin glass simulations
are very feasible. Physical quantities can be evaluated at all temperatures
from a single run. Tunneling through the high energy barriers and
sampling of independent ground states become possible, which certainly
are relevant enrichments
for the simulation of spin glasses. An investigation of the total weight
and the overlap pattern of the spin glass order parameter
might yield invaluable
information about their groundstate structure [13].\\
\\
{\Large \bf Acknowledgements}
\\

This work was partially supported by U.S. Department of Energy under
contracts DE-FG05-87ER40319 and DE-FC05-85ER2500, Deutsche
Forschungsgemeinschaft under contract H180411-1 and TUBITAK
of Turkey. TC and UH acknowledge the warm hospitality at SCRI.

\end{document}